\documentclass[english,pre,reprint,showpacs,superscriptaddress]{revtex4-1}
\usepackage[latin9]{inputenc}
\usepackage{color}
\usepackage{babel}
\usepackage{varioref}
\usepackage{prettyref}
\usepackage{amsmath}
\usepackage{amssymb}
\usepackage{graphicx}
\usepackage[unicode=true,
 bookmarks=true,bookmarksnumbered=false,bookmarksopen=false,
 breaklinks=true,pdfborder={0 0 0},backref=false,colorlinks=true]
 {hyperref}
\hypersetup{pdftitle={Event-Driven Simulations of a Plastic, Spiking Neural Network},
 pdfauthor={Chun-Chung Chen, David Jasnow}}
\usepackage{breakurl}

\makeatletter

\providecommand{\tabularnewline}{\\}

\@ifundefined{textcolor}{}
{%
 \definecolor{BLACK}{gray}{0}
 \definecolor{WHITE}{gray}{1}
 \definecolor{RED}{rgb}{1,0,0}
 \definecolor{GREEN}{rgb}{0,1,0}
 \definecolor{BLUE}{rgb}{0,0,1}
 \definecolor{CYAN}{cmyk}{1,0,0,0}
 \definecolor{MAGENTA}{cmyk}{0,1,0,0}
 \definecolor{YELLOW}{cmyk}{0,0,1,0}
 }

\makeatother

\begin{document}

\title{Event-driven simulations of a plastic, spiking neural network}

\author{Chun-Chung Chen}

\affiliation{Physics Division, National Center for Theoretical Sciences, Hsinchu,
Taiwan 300, Republic of China}

\affiliation{Department of Physics and Astronomy, University of Pittsburgh, Pittsburgh,
Pennsylvania 15260, USA}

\author{David Jasnow}

\affiliation{Department of Physics and Astronomy, University of Pittsburgh, Pittsburgh,
Pennsylvania 15260, USA}
\begin{abstract}
We consider a fully-connected network of leaky integrate-and-fire
neurons with spike-timing-dependent plasticity. The plasticity is
controlled by a parameter representing the expected weight of a synapse
between neurons that are firing randomly with the same mean frequency.
For low values of the plasticity parameter, the activities of the
system are dominated by noise, while large values of the plasticity
parameter lead to self-sustaining activity in the network. We perform
event-driven simulations on finite-size networks with up to 128 neurons
to find the stationary synaptic weight conformations for different
values of the plasticity parameter. In both the low and high activity
regimes, the synaptic weights are narrowly distributed around the
plasticity parameter value consistent with the predictions of mean-field
theory. However, the distribution broadens in the transition region
between the two regimes, representing emergent network structures.
Using a pseudophysical approach for visualization, we show that the
emergent structures are of {}``path'' or {}``hub'' type, observed
at different values of the plasticity parameter in the transition
region.
\end{abstract}


\maketitle

\section{Introduction}

Neurons can form plastic networks through connecting synapses with
weights that are changed dynamically by the neural activity in the
form of neural spikes. It has been established that the precise timing
of these spikes determines whether and how the synaptic weights will
be increased (potentiated) or decreased (depressed) \cite{abbott_synaptic_2000,van_rossum_stable_2000,roberts_spike_2002,dan_spike_2004}.
With better understanding of spike-timing-dependent plasticity (STDP),
it becomes increasingly important to find out its implications on
the underlying network structure and, consequently, on the neural
activity itself. Theoretical models of a plastic neural network typically
consist of three components: neural dynamics, synaptic transmission,
and network plasticity \cite{morrison_spike-timing-dependent_2007,chen_mean-field_2010}.
As has been shown previously \cite{chen_mean-field_2010}, a simple
self-consistent mean-field scheme can be constructed when these three
ingredients of a plastic neural network are given. Reducing the neural
and synaptic dynamics to a determination of response functions characterizing
a mean-field medium, the network plasticity is left governed by one-dimensional
random-walk dynamics. Such a mean-field scheme was applied \cite{chen_mean-field_2010}
to a network of leaky integrate-and-fire neurons \cite{gerstner_spiking_2002,burkitt_review_2006,brette_exact_2006},
coupled through a jump-and-decay synaptic conductance dynamics \cite{tsodyks_synchrony_2000},
and under STDP rules \cite{van_rossum_stable_2000}, controlled by
a \emph{plasticity parameter} $w^{\star}$ representing the expected
value of synaptic weights when firings of the neurons are purely driven
by noise. While we do not expect that there is a broad range of cell
types with different plasticity parameter values but otherwise similar
physiological characteristics, different values of $w^{\star}$ can
correspond to different stages of a (quasi-statically) growing and
developing network. The mean-field theory (MFT) predicts a first-order
phase transition and hysteresis from a low $w^{\star}$ regime of
noise-driven activity to a high $w^{\star}$ regime of self-sustaining
activity. It also predicts a narrow synaptic weight distribution as
long as the overall rate of change of synaptic strength is low. In
the current study, we perform intensive event-driven simulations \cite{brette_simulation_2007}
on the same model network to compare with the predictions of MFT;
the simulations also reveal emergent network dynamics and structures
that are not captured by the MFT.

In \prettyref{sec:Model} that follows, we briefly describe the simulated
model of a plastic neural network, which was elaborated in our MFT
study \cite{chen_mean-field_2010}. We then present in \prettyref{sec:Simulation-Method}
the method and algorithm of the event-driven simulations used to explore
the dynamics of the model. The results of the simulations for static
uniform networks are reported in \prettyref{sec:Static network},
where we present the mean activity levels for networks of different
sizes as functions of synaptic input. These activity levels agree
well with the MFT predictions in the stable phases of low and high
activity but show a gradual increase in a transition region ($w_{1}^{\star}<w^{\star}<w_{2}^{\star}$,
from the onset to the persistence of threshold firing events) instead
of the sharp first-order jump predicted by MFT. We also analyze the
sporadic patterns of self-sustaining threshold firings in the transition
region to identify two types of threshold firing events. In \prettyref{sec:Dynamic-network},
we present simulation results with high resolution in the plasticity
parameter $w^{\star}$ for a plastic network with $N=32$ neurons
that has reached a stationary state of plasticity. The change in the
synaptic-weight conformation of the network in the transition region
manifests itself in the synaptic weight distribution, which is seen
to broaden twice, along with a bimodal elevation of the average firing
activity compared to that of a static network. To characterize the
emergent synaptic-weight structure of the network in the transition
region, we employ a pseudophysical approach for visualization in \prettyref{sec:Network-Layout}
to generate a 2D layout of the network. Such an approach reveals a
\emph{path} or \emph{loop} conformation near the lower end of the
transition region ($w^{\star}\sim w_{1})$ for network sizes up to
$N\approx32$ and a \emph{hub} conformation near the high end ($w^{\star}\sim w_{2}$)
for $N\approx24$ and greater. We then conclude and summarize our
findings in \prettyref{sec:Conclusions}.

\section{Model\label{sec:Model}}

While our method is applicable to other combinations of models of
neurons, synapses and plasticity, we follow the same choices made
in our mean-field approach \cite{chen_mean-field_2010}, which we
describe briefly below.

In the integrate-and-fire model for the neurons, the state of a neuron
$i$ is described by a membrane potential $V_{i}$, which follows
the differential equation of a leaky integrator \cite{dayan_theoretical_2001}

\begin{equation}
\tau_{m}\frac{dV_{i}}{dt}=V_{0}-V_{i}+G_{i}\left(\mathcal{R}-V_{i}\right),\label{eq:membrane-potential}
\end{equation}
where $\tau_{m}$ is the leak time for the membrane charge, $V_{0}$
is the resting potential when the neuron is in the quiescent state,
and $\mathcal{R}$ is the reversal potential for the ion channels
on the synapses. The total synaptic conductance $G_{i}$ for the neuron
is given by the sum 
\begin{equation}
G_{i}\equiv\sum_{j}w_{j,i}Y_{j}\label{eq:total-conductance}
\end{equation}
over all presynaptic neurons, $j$. The synaptic weights $w_{j,i}$
define the network and the same active transmitter fraction $Y_{j}$
is assumed for all efferent synapses of neuron $j$. In addition to
the continuous dynamics \eqref{eq:membrane-potential}, a neuron fires
when its membrane potential reaches a threshold value, $V_{\mathrm{th}}$.
Then, its membrane potential drops immediately to a reset value, $V_{\mathrm{r}}$.
The action potential of the integrate-and-fire model is assumed to
be instantaneous and is not modeled explicitly. The spike train produced
by the neuron $i$ is defined as the function 
\begin{equation}
S_{i}\left(t\right)\equiv\sum_{n}\delta\left(t-t_{i,n}\right)\label{eq:spike-train}
\end{equation}
 where $t_{i,n}$ is the time when the neuron $i$ fires for the $n$-th
time. 

The fraction $Y_{j}$ of the active transmitters is described by the
Tsodyks--Uziel--Markram (TUM) model \cite{tsodyks_synchrony_2000}
of neural transmission, where the transmitters are distributed in
three states: {}``active'', with the fraction $Y$; {}``inactive'',
with the fraction $Z$; and {}``ready-to-release'', with the fraction
$X$. For efferent synapses of a presynaptic neuron $j$, these fractions
follow the dynamics \cite{tsodyks_synchrony_2000} 
\begin{eqnarray}
\frac{dX_{j}}{dt} & = & \frac{Z_{j}}{\tau_{R}}-uS_{j}X_{j}\nonumber \\
\frac{dY_{j}}{dt} & = & -\frac{Y_{j}}{\tau_{D}}+uS_{j}X_{j}\label{eq:TUM-model}\\
\frac{dZ_{j}}{dt} & = & \frac{Y_{j}}{\tau_{D}}-\frac{Z_{j}}{\tau_{R}},\nonumber 
\end{eqnarray}
where $\tau_{D}$ is the decay time of active transmitters to the
inactive state, $\tau_{R}$ is the recovery time for the inactive
transmitters to the ready-to-release state, and $u$ is the fraction
of ready-to-release transmitters that is released to the active state
by each presynaptic spike. With the conservation rule 
\begin{equation}
X_{j}+Y_{j}+Z_{j}=1,\label{eq:transmitter-conservation}
\end{equation}
there are two independent variables per presynaptic neuron. Consistent
with the TUM dynamics, the values of the factors multiplying $S_{j}$
at the discontinuities are to be evaluated immediately before the
discontinuities.

While the integrate-and-fire and TUM dynamics are both deterministic,
we model the stochasticity of the network with additional noise-driven
firing events following Poisson statistics with the frequency $\lambda_{N}$
for each neuron. The noise-driven firings are treated the same way
as threshold firings, that is, the membrane potentials are brought
instantaneously to the reset value $V_{\mathrm{r}}$ and the firing
times are included in the spike trains \eqref{eq:spike-train} of
the transmitter dynamics \eqref{eq:TUM-model}.

To minimize the computational cost, the active transmitter fractions
$Y_{j}$ double as the exponentially decaying {}``window functions''
in our version of the plasticity rules suggested by van Rossum \emph{et
al.} \cite{van_rossum_stable_2000}. The synaptic weights are taken
to follow the dynamical equation 
\begin{equation}
\frac{dw_{j,i}}{dt}=\Delta Y_{j}S_{i}-rw_{j,i}Y_{i}S_{j}\label{eq:simplified stdp}
\end{equation}
where $\Delta$ is the parameter for additive potentiation and $r$
is the parameter for multiplicative depression. We define the \emph{plasticity
parameter} as the ratio 
\begin{equation}
w^{\star}\equiv\frac{\Delta}{r},\label{eq:wstar-definition}
\end{equation}
which is equal to the expectation value of the synaptic weight when
we have the symmetry $\left\langle Y_{j}S_{i}\right\rangle =\left\langle Y_{i}S_{j}\right\rangle $,
\emph{e.g.}, when $S_{i}$ and $S_{j}$ are both Poisson spike trains
of the same frequency. Fixing $w^{\star}$ leaves the depression factor
$r$ as an overall control parameter for the \emph{rate of plasticity}.

\begin{table}
\caption{Values of parameters used in calculations\label{tab:Values-of-parameters}}

\centering{}\begin{tabular}{rlccrl}
\multicolumn{2}{c}{integrate and fire} &  &  & \multicolumn{2}{c}{TUM model}\tabularnewline
\cline{1-2} \cline{5-6} 
resting potential $V_{0}$: & -55 mV &  &  & decay time $\tau_{D}$: & 20 ms\tabularnewline
leak time $\tau_{m}$: & 20 ms &  &  & recovery time $\tau_{R}$: & 200 ms\tabularnewline
firing threshold $V_{\mathrm{th}}$: & -54 mV &  &  & release fraction $u$: & 0.5\tabularnewline
\cline{5-6} 
reset potential $V_{\mathrm{r}}$: & -80 mV &  &  & noise frequency $\lambda_{N}$: & 1 Hz\tabularnewline
reversal potential $\mathcal{R}$: & 0 mV &  &  & plasticity rate $r$: & 0.01\tabularnewline
\end{tabular}
\end{table}
It is common for a model of a biological system to carry a large number
of empirical parameters. Instead of exploring all possible ranges
of these parameters, we fix them with physiologically plausible values
that are commonly found in the literature. Unless otherwise stated,
the values of the parameters are as listed in Table \ref{tab:Values-of-parameters}.

\section{Simulation Method\label{sec:Simulation-Method}}

The disparity between the time scales of spiking activities and neural
plasticity posts a significant challenge to computer simulations of
plastic neural networks. Particularly for STDP, where precise timing
is crucial in determining synaptic changes, we can not alleviate the
computation requirement through use of larger integration time steps.
However, for continuous dynamics that can be solved analytically,
one can improve the efficiency of computation through an \emph{event-driven}
approach similar to that described by Brette \cite{brette_exact_2006},
which gives machine precision timing for the spikes and requires limited
amount of computation upon each spike production.

In the event-driven approach, instead of calculating the state of
the system for a fixed increment of time, one calculates the time
for the next discontinuous event. This approach is feasible when the
continuous dynamics of the system is solvable so that the time for
the next discontinuous event can be evaluated efficiently. For the
leaky integrate-and-fire model considered, an analytical form can
be written down for the trajectory of the normalized membrane potential
\cite{brette_exact_2006}
\begin{equation}
v_{i}\equiv\frac{V_{i}-V_{0}}{\mathcal{R}-V_{0}}=xe^{x}\left[E_{1}\left(x\right)-E_{1}\left(G\right)+\frac{v}{Ge^{G}}\right]\label{eq:IF trajectory}
\end{equation}
where $E_{1}$ is the exponential integral function of the first kind,
$x\equiv Ge^{-t/\tau_{m}}$, and $v=v_{i}\left(0\right)$, $G=G_{i}\left(0\right)$
are the initial states of the neuron. For the TUM dynamics, the transmitter
fractions follow the trajectories 
\begin{eqnarray}
Y_{i} & = & e^{-t/\tau_{D}}Y\nonumber \\
Z_{i} & = & \left[Z+\frac{\tau_{R}}{\tau_{R}-\tau_{D}}Y\right]e^{-t/\tau_{R}}-\frac{\tau_{R}}{\tau_{R}-\tau_{D}}Y_{i}\label{eq:transmitter trajectory}
\end{eqnarray}
where $Y=Y_{i}\left(0\right)$ and $Z=Z_{i}\left(0\right)$ are their
initial values. With the trajectory \eqref{eq:IF trajectory}, one
can solve for the time of the next threshold crossing $V_{i}\rightarrow V_{\mathrm{th}}$
through, \emph{e.g.}, a root solver. The computation time for a root
solver to find the position of the crossing depends only logarithmically
on the precision required, compared to the linear increase for the
fixed-time-step approach. Further improvement of the computational
efficiency can be achieved by, \emph{e.g.}, precomputing a lookup
table for the time to threshold crossing given the current state of
a neuron and interpolating when the desired value falls between precomputed
points. For the simulated network with Poisson noise, the algorithm
is as outlined below:
\begin{enumerate}
\item For each neuron $i$ in the system, calculate the time-to-fire $t_{\mathrm{ttf}}^{i}$
from the continuous, deterministic dynamics of the model.\label{enu:For-each-neuron}
\item Draw the time to its next noise-triggered firing $t_{N}^{i}$ from
an exponential distribution with the expectation value $\lambda_{N}^{-1}$
for each neuron $i$.
\item Find the minimum of the times $t_{\mathrm{ttf}}^{i}$ and $t_{N}^{i}$
among all neurons to be the time to the next discrete event $\Delta t$
of the system.
\item Advance the simulated clock by $\Delta t$ and update the state of
the system using Eqs. \eqref{eq:IF trajectory} and \eqref{eq:transmitter trajectory}.
\item Fire the selected neuron $i$ by reseting its membrane potential to
$V_{i}=V_{R}$, and increasing the affected active transmitter fraction
$Y_{i}$ by $u\left(1-Y_{i}-Z_{i}\right)$.
\item Repeat this process from \prettyref{enu:For-each-neuron}.
\end{enumerate}
For the current model, the simulated dynamics boils down to the evaluation
of the time-to-fire $t_{\mathrm{ttf}}\left(V,G\right)$ given the
initial membrane potential $V$ and total synaptic conductance $G$
as well as the evaluation of the trajectories \eqref{eq:IF trajectory}
and \eqref{eq:transmitter trajectory}. We note that instead of the
common approach of defining the model with the set of differential
equations \eqref{eq:membrane-potential} and \eqref{eq:TUM-model},
it is equally valid and computationally preferable to define the continuous
dynamics of the model with the trajectories \eqref{eq:IF trajectory}
and \eqref{eq:transmitter trajectory} for the neurons and synapses.

\section{Static network\label{sec:Static network}}

\begin{figure}
\begin{centering}
\includegraphics[width=1\columnwidth]{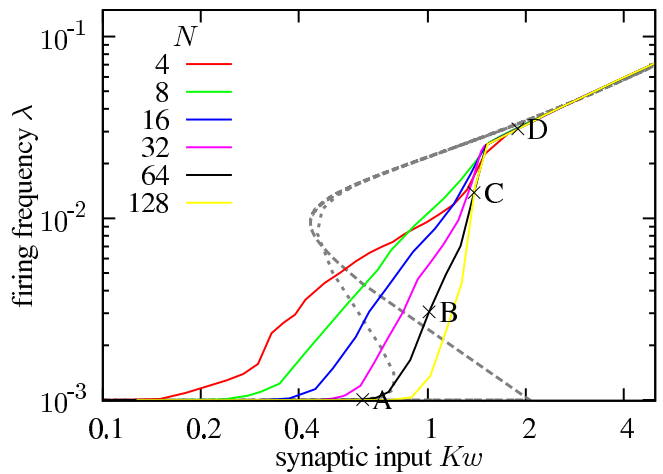}
\par\end{centering}

\caption{(Color online) Mean activity level vs. total synaptic input ($Kw$)
for fully-connected uniform networks (node degree $K=N-1$) with static
synaptic weight ($w$). Dashed lines are predictions from mean-field
theory without fluctuation corrections (long dashes) and with the
inclusion of shot-noise-like fluctuations (short dashes) in total
synaptic conductance for a network with $K=63$ afferent synapses
per neuron. \label{fig:static network activity}}
\end{figure}

Before incorporating plasticity, we first perform event-driven simulations
on fully-connected networks of up to 128 neurons with uniform, fixed
synaptic weights $w$. This exercise will prepare us for the inclusion
of plasticity and also reveal the emergence of {}``structure'' in
the network. The average activity levels for different system sizes
are shown in Fig.~\ref{fig:static network activity} 
compared with predictions from MFT \cite{chen_mean-field_2010}. As
expected, the average activity levels of the simulated network coincide
with mean-field predictions in the stable phases of low and high activity.
However, in the transition region between the two phases, the activity
levels from network simulations increase continuously instead of exhibiting
the jumps and hysteresis predicted by mean-field theory. We note that
the hysteresis in MFT \cite{chen_mean-field_2010} comes about when
there are two stable states in the system: a quiet state with only
noise-triggered firings that are insufficient to ignite the entire
system, and an active state in which firings are system-wide and self-sustaining.
However, for a finite-size network having sufficient fluctuations
and running for a sufficient time, the system can make transitions
between the two states, leading to a single-valued mean activity level
of the system showing no hysteresis. Comparing networks of different
sizes, the rise of the average firing frequency in the transition
region is steeper for larger networks since fluctuations in the total
synaptic conductance \emph{decrease} with an \emph{increase} in the
number of afferent synapses \cite{chen_mean-field_2010} rendering
mean-field theory a better approximation.

\begin{figure}
\begin{centering}
\includegraphics[width=1\columnwidth]{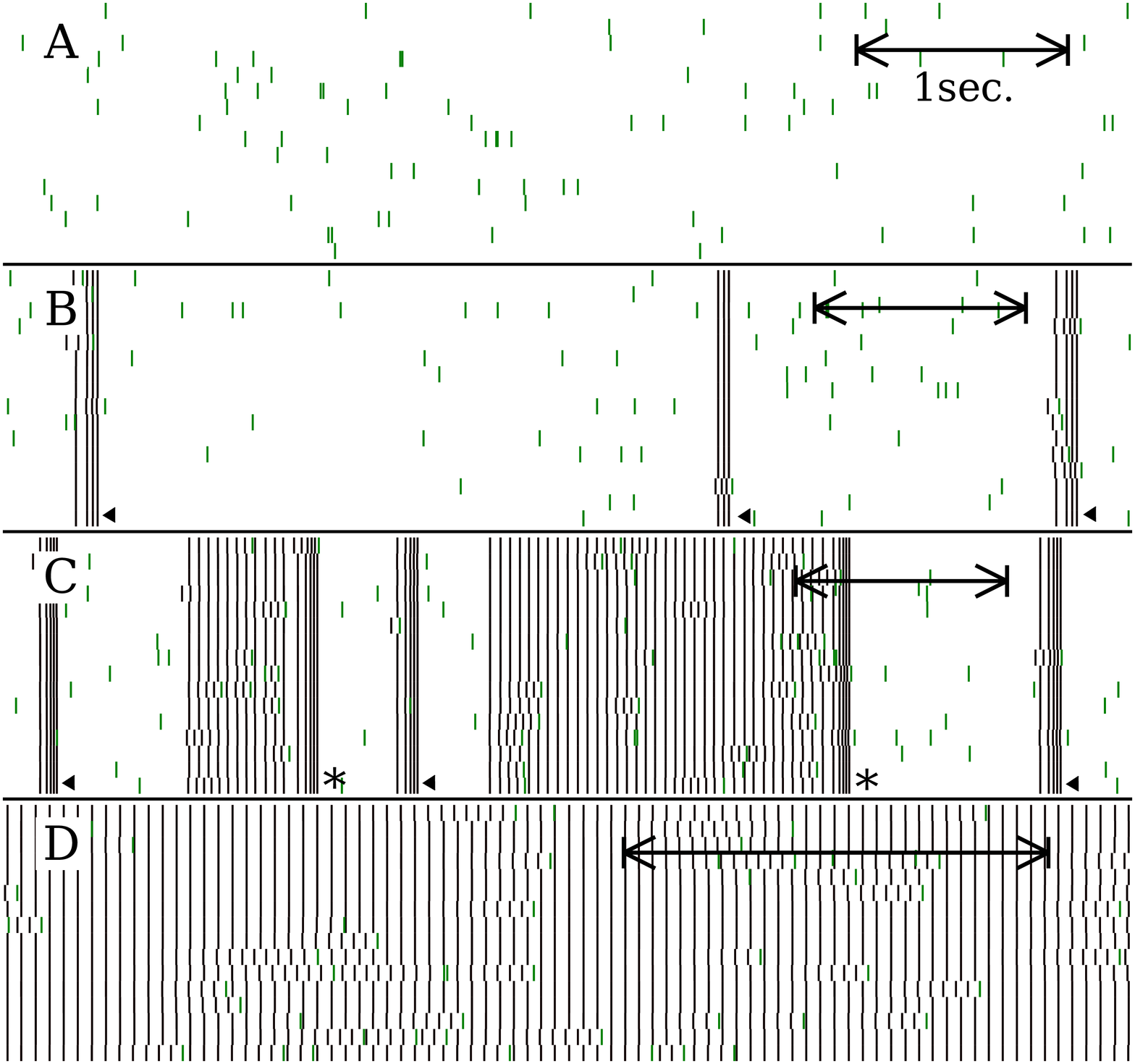}
\par\end{centering}

\caption{(Color online) Firing polygraphs of the $N=64$ network at 4 different
values of synaptic strength $w$ labeled in Fig.~\ref{fig:static network activity}.
For each value, only activities of 16 randomly chosen neurons are
shown, with each occupying 1/16 of the height of the corresponding
strip. The light and dark vertical line segments mark the firings
due to noise and threshold crossing, respectively. The time on each
strip of recording runs towards the left with a scale indicated by
the one second time interval on each strip. Symbols ($\blacktriangleleft$
and $\ast$) mark clusters of threshold firings as detailed in text.\label{fig:firing polygraphs static}}
\end{figure}

\begin{figure}
\begin{centering}
\includegraphics[width=1\columnwidth]{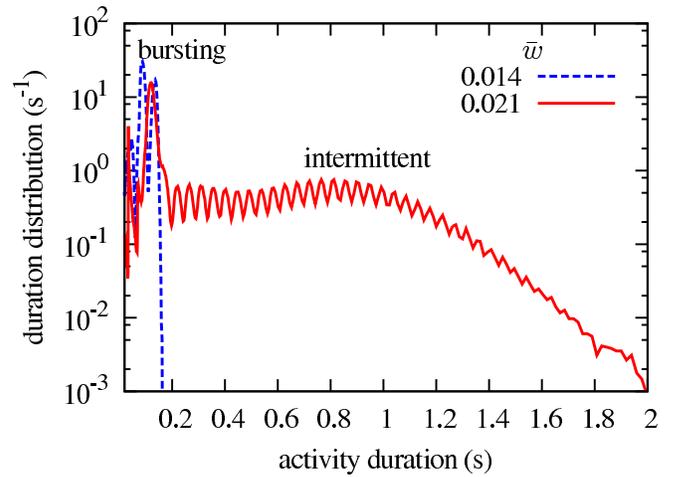}
\par\end{centering}

\caption{(Color online) Time-weighted duration distribution of sustaining threshold
firing activities. The system is considered active when any of the
neurons is set to cross the firing threshold without further noise
events. The duration is measured for each episode of the system staying
continuously active. The oscillations in the distribution comes from
the periodic nature of persistent firing activities. Values of the
synaptic weight $\bar{w}$ (0.014, dashed line and 0.021, solid line)
correspond to the locations B and C, respectively in Figs.~\ref{fig:static network activity}
and \ref{fig:firing polygraphs static}.\label{fig:Duration-distribution}}
\end{figure}

The firing pattern of the $N=64$ network is shown in Fig. \ref{fig:firing polygraphs static}
for 4 different values of synaptic input $Kw$ as marked in Fig.~\ref{fig:static network activity}.
 With computer simulations, we are able to distinguish firings due
to threshold crossings of the membrane potentials (marked with dark
vertical line segments) from noise triggered firings (marked with
light vertical line segments). We can thus define a system to be active
when there is any neuron with a membrane potential set to cross the
firing threshold without the help of further noise events. Operationally,
this is when the time-to-fire $t_{\mathrm{ttf}}^{i}$ calculated in
step \ref{enu:For-each-neuron} of the event-driven algorithm \vpageref{enu:For-each-neuron}
is finite for any of the neurons. The switching between the quiet
and active states of the network are evident on the strips B and C.
This leads to clusters of threshold firings during which the system
remains active. In the specific model we considered, there are two
types of threshold firings clusters that can be identified from the
polygraphs. The first represents \emph{bursting}, which are brief
clusters (up to a few hundred milliseconds) of threshold firings of
generally similar durations. These can be seen for all threshold activities
on strip B and some on strip C as marked by the symbol {}``$\blacktriangleleft$''
in Fig.~\ref{fig:firing polygraphs static}. The second type of threshold
firing activity is \emph{intermittently} \emph{persisting} activity,
that can last for seconds, are of a wider range of durations and are
seen in two time segments marked by the symbol {}``$\ast$'' on
strip C of Fig.~\ref{fig:firing polygraphs static}. The two types
of clusters of threshold firings can also be identified from the duration
distribution of self-sustained threshold firings shown in Fig.~\ref{fig:Duration-distribution}.
The distinction between the two kinds of activities can be attributed
to the short-term depression caused by the inactive state of synaptic
transmitters in the TUM model: When threshold firings of the network
are just ignited after a quiet period, most transmitters are in the
ready-to-release state and activities can propagate easily leading
to a somewhat higher firing rate of the neurons in the beginning.
However, as most transmitters are deposited into the inactive state
during repeated firings, the firing frequency is reduced. At the low
activity end of the transition region, the reduction of firing frequency
typically continues all the way to a cessation of any threshold firings,
leading to the short bursting events of similar duration. Near the
high activity end of the transition region, as the transmitters become
inactive, the reduced firing frequency can settle to a stable value
that can last for seconds before all threshold firings stop. 
\begin{figure}
\begin{centering}
\includegraphics[width=1\columnwidth]{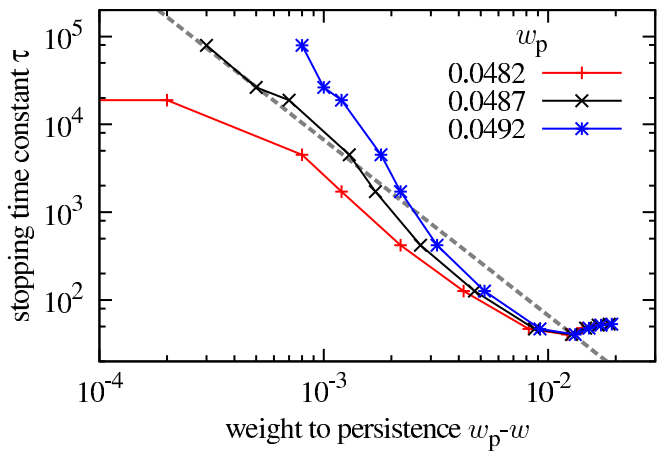}
\par\end{centering}

\caption{(Color online) Scaling of {}``stopping time constant'' estimated
from the tail of active duration distribution in Fig.~\ref{fig:Duration-distribution}
for a $N=32$ static uniform network. The best estimate of $w_{\mathrm{p}}=0.0487$
for the diverging point (center curve) is shown along with two slight
deviations. The dashed line is to show the estimated slope of -2 for
comparison. \label{fig:Scaling-of-stopping-time}}
\end{figure}
From the simulation results such as what are shown in Fig.~\ref{fig:Duration-distribution},
the stopping of the threshold firings seem to follow Poisson statistics,
that is, it can be characterized by a stopping rate $\tau^{-1}$.
The duration distribution of sustaining threshold firings decays exponentially
for large duration with the time constant $\tau$ that increases with
$w$ and appears to diverge at some $w=w_{\mathrm{p}}$. For the $N=32$
network, we estimate $w_{\mathrm{p}}\approx0.0487$ and that the {}``stopping
time constant'' diverges roughly with the scaling $\tau\sim\left(w_{\mathrm{p}}-w\right)^{-2}$
as suggested in Fig.~\ref{fig:Scaling-of-stopping-time}. We note
that the scaling form of the stopping time remains the same when we
simplify the TUM dynamics in our model by removing the inactive state
(data not included here), suggesting at least some degree of universality.
These results suggest that the divergence of the stopping time constant
$\tau$ signals that the system enters the phase of truly self-sustaining
activity.

\section{Dynamic (Plastic) network\label{sec:Dynamic-network}}

For the current study, we focus on the stationary states of the system
under the plasticity dynamics. A uniform network with synaptic weight
$w_{j,i}=w^{\star}$ for all synapses is used as the initial configuration,
and simulations are conducted at each data point for at least $2^{31}$
ms ($\approx25$ days) of simulated time to reach the stationary state.
We concentrate our calculations on a fully-connected network of $N=32$
neurons with a high resolution in the plasticity parameter $w^{\star}$.
For each $w^{\star}$, all measurements are averaged over an ensemble
of 32 or more independent runs of different random sequences. 
\begin{figure}
\begin{centering}
\includegraphics[width=1\columnwidth]{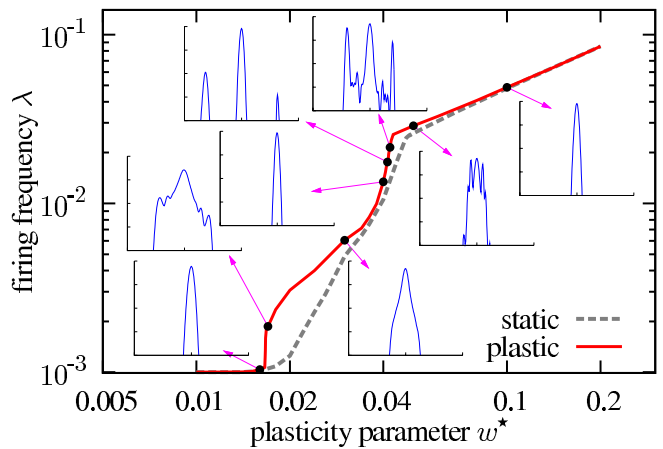}
\par\end{centering}

\caption{(Color online) Activity--plasticity(parameter) plot for a fully connected
plastic network of $N=32$ neurons compared with the activity level
of a static network of the same size with uniform weight $w=w^{\star}$.
The plastic network is allowed to reach a stationary state by running
for about 25 days of simulated time. All results are averaged over
an ensemble of 32 or more runs of different random sequences. Insets
of the plot show the normalized synaptic weight distributions on a
log-log scale with a range of 2 (4) decades on the horizontal (vertical)
axis at different points marked on the activity--plasticity curve.\label{fig:plastic activity}}
\end{figure}
The results of the calculations are summarized in Fig.~\ref{fig:plastic activity},
where we see that the firing frequency of a stationary \emph{plastic}
network coincides with that of a uniform \emph{static} network ($w=w^{\star}$
for all synapses) in the low-activity, noise-dominated regime as well
as in the persistently active regime. In the transition region between
the noise-driven and self-sustaining regimes, the firing rate of the
network is enhanced by the plasticity, and, while the behavior is
not universal, with TUM synapses the amount of enhancement exhibits
an interesting bimodal shape.

In the insets of Fig.~\ref{fig:plastic activity}, we show the synaptic
weight distribution, also averaged over the same ensemble, at each
marked point along the activity--plasticity curve. In the two stable
regimes, noise-dominated and persistently-active, the synaptic weights
have a Gaussian distribution with a narrow width (within about 5\%
of $w^{\star}$) as predicted by MFT \cite{chen_mean-field_2010}.
However, coincident with the bimodal enhancement of firing frequency,
the synaptic weight distribution of the stationary plastic network
shows dramatic changes in the transition region along the activity--plasticity
curve: Increasing the plasticity parameter $w^{\star}$ from the noise-dominated
low-activity regime, we see a discontinuous jump in the firing rate
and sudden broadening of the weight distribution with tiered side
peaks at $w^{\star}\approx0.017$. The enhanced firing and broadened
distribution ease off after the jump, until these features become
insignificant around $w^{\star}=0.04$. At this point, with a slight
increase of $w^{\star}$, two disconnected side peaks pop out in the
synaptic weight distribution. However, such a splitting in the weight
distribution is only accompanied by a gradual enhancement of average
firing rate of the plastic network. The two side peaks eventually
merge back to the main peak with further increases of $w^{\star}$.
This eventually returns the network to a uniform conformation with
a narrow Gaussian synaptic-weight distribution before the plasticity
parameter reaches $w^{\star}\simeq0.1$.

Instead of a simple increase of the Gaussian width, the observed broadening
of the synaptic weight distribution in the transition region just
summarized (see Fig.~\ref{fig:plastic activity}) comes with structured
side peaks, which signify emergent network conformations that we try
to decipher and visualize with a pseudophysical approach in \prettyref{sec:Network-Layout}.
For the $N=32$ and smaller networks, we are able to obtain a stationary
state from the simulation run that is insensitive to the initial configuration
at each data point (value of $w^{*}$). However, we do see runaway
synaptic weights under the plasticity rules used for the $N=48$ and
$N=64$ networks in narrow, isolated intervals at $w^{\star}\approx0.0268$
and $w^{\star}\approx0.0194$ respectively near the high-activity
end of the transition region before the runs ($2^{31}$ms, $\sim$25
days, of simulated time) are complete. For these instances, the diverging
synaptic weights and firing frequencies of the driven neurons slow
the simulation down to a near standstill, and we can not always maintain
a stationary system as we vary the plasticity parameter across the
runaway point. We note that such runaways are commonly seen in models
with Hebbian plasticity and are typically {}``cured'' with cutoffs
in the range of synaptic weights. \cite{amarasingham_predictingdistribution_1998,song_competitive_2000,rubin_equilibrium_2001,toyoizumi_optimality_2007}
We do not need such a device for the $N=32$ network results presented.
For larger networks, when we do see such {}``isolated'' runaways,
we regard them as pathological \cite{hasselmo_runaway_1994,greenstein-messica_synaptic_2011}
and exclude them from the ensembles.

\section{Network {}``Layout''\label{sec:Network-Layout}}

\begin{figure*}
\centering{}\includegraphics[width=0.9\textwidth]{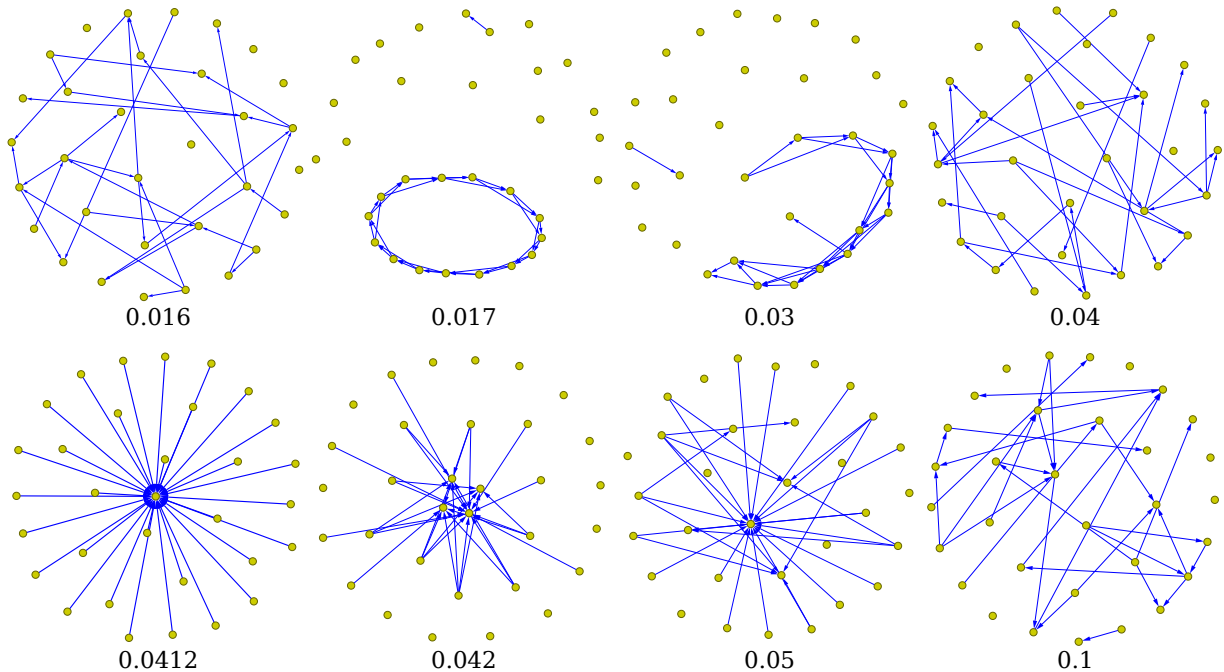}\caption{(Color online) Typical resultant layouts of neurons (represented by
the dots) following the pseudophysical approach described in the text
at the points marked on the activity--connectivity curve in Fig.~\ref{fig:plastic activity}.
The number labeled below each layout is the corresponding value of
the plasticity parameter $w^{\star}$. Only the strongest 31 synaptic
connections (lines with arrows) are shown in each layout.\label{fig:layouts}}
\end{figure*}
Characterizing the structure of non-uniform networks is an active
field of research \cite{bullmore_complex_2009,boccaletti_complex_2006,newman_modularity_2006}.
Here we adopt a more intuitive and visual approach to extract any
emergent structures of the stationary plastic network obtained above.
Under our model setup, such structures reside in the framework of
a fully-connected network of uniform connectivity, represented by
the main peak in the synaptic weight distribution. In this section,
we attempt to uncover the topological meaning of the side peaks seen
for the distribution in the transition region between the two stable
phases of the system. 

In order to visualize the networks such that any emerging features
can be better revealed, we introduce a pseudophysical system of fictitious
interacting particles in two dimensions representing the neurons,
recognizing, of course, that the actual all-to-all network is not
strictly two-dimensional. The nodes (pseudo particles) are imagined
to be located on the plane, confined to the region near an arbitrary
origin by an isotropic harmonic potential. The nodes interact with
a (fictitious) pair-wise repulsive force that is determined by the
synaptic weights between the neurons. To have the pseudo-particles
representing neurons that are more strongly connected stay close to
one another, we choose a repulsive pairwise force 
\begin{equation}
\vec{F}_{i,j}=\frac{\hat{e}_{j,i}}{w_{i,j}^{2}+w_{j,i}^{2}}
\end{equation}
to act on particle $i$ due to $j$ with $\hat{e}_{j,i}=\vec{r}_{j,i}/\left|\vec{r}_{j,i}\right|$
being the unit vector along the $\vec{r}_{j,i}=\vec{r}_{i}-\vec{r}_{j}$
direction (in the two-dimensional pseudo-space). In fitting the entire
layout of the pseudo-particles (representing neurons) to a fixed display
area, the strength $k$ of the confining harmonic potential is adjusted
so that, in an equilibrium arrangement of the particles, the radial
coordinate $\vec{r}$ of the particle farthest from the origin of
the potential has magnitude equal to $1$. Such a pseudophysical system
of neurons is relaxed to reach a stable arrangement of neurons in
which the net force on each pseudo particle (neuron) is zero. The
relaxation is accomplished through a deterministic, over-damped dynamics
in which the velocity of each node is proportional to the net force
it experiences
\begin{equation}
\frac{d\vec{r}_{i}}{dt}=-k\vec{r}_{i}+\sum_{j}\vec{F}_{i,j}.
\end{equation}

The representative layouts for the marked data points in Fig.~\ref{fig:plastic activity}
are shown in Fig.~\ref{fig:layouts}, where we only show a fraction
of the synapses with strongest weights along with the pseudo-positions
to reduce obscurity. We note that the stable layouts reached are not
unique for the system, and it is possible for the pseudo-particle
system to be trapped in a metastable configuration in which strongly
connected neurons are hindered by other neurons and fail to get close
to each other during the relaxation. We do not attempt to further
relax the layout (by, \emph{e.g.}, perturbing the positions of the
neurons) after a stable configuration is reached during the relaxation
process. (We note that for larger systems where the {}``hindrance''
can be more severe, one may consider a three or higher dimensional
pseudophysical system to help alleviate the problem.)

As revealed in layouts of Fig.~\ref{fig:layouts}, the jump in the
firing rate near $w^{\star}=0.017$ is accompanied by the formation
of a \emph{path} conformation in the network where synapses of stronger
synaptic weights connect a sequence of neurons and facilitate the
propagation of activities along the path. Such a path can form a closed
\emph{loop} near the jump but become less likely to do so as $w^{\star}$
is increased. The path or loop structure is stable in the system such
that once it forms, it will generally remain to the end of our simulation
run at a fixed $w^{\star}$ value. As we only present results of simulations
from uniform initial networks, there is no hysteresis loop in Fig.~\ref{fig:plastic activity}.
However, the jump in the firing rate signals a first-order phase transition.
While we have not attempted a systematic study of hysteresis effects
with larger systems, we do see some degree of hysteresis in smaller
($N\leq16$) networks: If we use a loop conformation obtained in the
stationary state of the network at a $w^{\star}$ value above the
jump as the initial condition for the simulation of the network at
a $w^{\star}$ value below the jump, the loop conformation can sometimes
persist over the course of the entire simulation. With a path or loop
conformation of the network, the tiered side peaks of the synaptic
weight distribution in this area come from the skip-over synapses
along the path,\emph{ i.e.}, synapses that are along the direction
of the path but bypass some of the neurons.

Eventually, with the increase of $w^{\star}$, the path falls apart
and the network returns to a uniform conformation around $w^{\star}=0.04$.
The emergence of disjoint side peaks in the synaptic weight distribution
with a slight increase of $w^{\star}$ from $0.04$ as seen in Fig.~\ref{fig:plastic activity}
is accompanied by the formation of a \emph{sink hub} structure in
the network where the high activity level of the hub neuron and its
strong afferent synapses reinforce each other because of the positive-feedback
nature of the synaptic plasticity. This type of hub structure is first
seen at $w^{\star}=0.0411$ in our simulations and can go in and out
of existence several times during the course of our simulation run
at a fixed $w^{\star}$. This partially explains why we only see a
gradual increase of the average activity level when a hub start to
form. The sink hub structure becomes more and more stable as $w^{\star}$
increases until multiple hubs appear in the network around $w^{\star}=0.0414$.
The network returns to a uniform conformation through the increasing
number of less and less significant hubs. Finally, at sufficiently
large $w^{\star}$ the network has reached a homogeneous, high-activity
state.

We can now try to understand the formation of {}``path'' and {}``hub''
conformations in different segments of the transition region separating
homogeneous, stable low- and high-activity states of the network.
Near the low-activity end of the transition region, the sequence of
neurons along the path represents a firing order in a bout of threshold
firings that is amplified and preserved by the STDP as the path of
stronger synaptic weights. Especially when such a path forms a loop,
threshold firing activity can cycle through all neurons in the loop
no matter which one of them is triggered by the noise. However, near
the high-activity regime, each neuron generally fires multiple times
during an episode of sustaining threshold firings with a high frequency.
The firing order of neurons loses its meaning because there are many
ways of pairing up their spikes. In this area of the transition region,
the main distinguishable feature of a neuron is simply its firing
rate. In fact, the expectation value of a synaptic weight $w$ for
a synapse in this range can be uniquely determined from the firing
rates of its pre- and postsynaptic neurons when the firing rates of
the neurons are high enough ($\lambda>\tau_{D}^{-1},\tau_{m}^{-1}$)
so that the actual timing of spikes becomes unimportant. Under such
a condition, the stationary synaptic weight under the plasticity dynamics
\prettyref{eq:simplified stdp} for, say, the $j\rightarrow i$ synapse,
is given by 
\begin{equation}
w_{j,i}=w^{\star}\frac{\bar{Y}_{j}\lambda_{i}}{\bar{Y}_{i}\lambda_{j}}\label{eq:active-phase-weight}
\end{equation}
where $\bar{Y}_{i}\equiv\left\langle Y_{i}\right\rangle $ and $\lambda_{i}\equiv\left\langle S_{i}\right\rangle $
are, respectively, the average active transmitter fraction and mean
firing rate of the neuron $i.$ With Eq.~\eqref{eq:active-phase-weight},
the condition 
\begin{equation}
w_{i,j}w_{j,k}w_{k,i}=w^{\star3}\label{eq:triplet-weight}
\end{equation}
should hold for any neurons $i\neq j\neq k$ in the network %
\footnote{The loop product $w_{i,j}w_{j,k}w_{k,i}$ is an attempt to generalize
the curl operator $\nabla\times$ in metric spaces to a network. The
condition \eqref{eq:triplet-weight} is equivalent to the curl-free
condition of the {}``vector field'' $v_{i,j}\equiv\ln\left(w_{i,j}/w^{\star}\right)$,
which seems to hold in the hub conformations but not the loop or path
conformations for the stationary plastic networks.%
}. And, indeed, we have verified that the condition \eqref{eq:triplet-weight}
is well satisfied by the hub conformations obtained in the simulations,
but is apparently violated by the path conformations (data analysis
not presented here).

While it is tempting to correlate the layouts resulting from our pseudophysical
arrangement with configurations of physical neurons, we must caution
that one can only view such a pseudophysical approach as an arbitrary
and non-unique way of revealing network structures. Furthermore, the
structures found by the pseudophysical approach do not represent any
direct physical information contained in the parameters of the neuron
network. Nonetheless, it helps to provide an intuitive {}``structural''
understanding of the network conformation as we have shown above.

\section{Conclusions\label{sec:Conclusions}}

The event-driven algorithm presented in Sec.~\ref{sec:Simulation-Method}
improves the accuracy and efficiency of the simulations allowing us
to gain an intensive view of the phase space of a stationary plastic
neural network. With a large number of parameters typical for a model
of a biological system, we fix all but one, the plasticity parameter,
with physiologically plausible values for our investigation. Also,
to minimize conceptual complications, we study only the stationary
properties of fully-connected networks driven by uniform, non-correlated
Poisson noise. Our results show the network develops interesting structures
in the transition region between the stable phases corresponding to
low- and high-activity regimes. While our fully-connected network
with a limited number of neurons is certainly inadequate to address
the dynamics of a real brain, it can be a reasonable starting point
for cultured networks consisting of hundreds of neurons with virtually
{}``all-to-all'' interactions \cite{bi_synaptic_1998,segev_observations_2001,shefi_morphological_2002,beggs_neuronal_2003,lai_growth_2006}.
Our finding suggests that emergent structures in these networks are
more likely to be seen when it is firing intermittently in the transition
region perhaps during its development from a weakly-coupled system
of neurons to a more strongly connected network.

The characterization of network connectivity is an active field of
research with established quantifications such as the {}``clustering
coefficient'' \cite{barrat_architecture_2004} and {}``modularity''
\cite{newman_modularity_2006}. However, in the current study, we
adopt a simple visualization approach to gain a more intuitive view
of the network structure. The resultant identification of the {}``path''
and {}``hub'' conformations explains or, more accurately, coincides,
with the appearance of structured side peaks in the synaptic weight
distribution and the elevated firing rate of the neurons. While these
path and hub conformations are the simplest forms that can appear
in a connected network, the present work is the first to demonstrate
that they can arise naturally in transition regions of a neural network
under STDP and driven only by noise. The connection-weight conformation
(\emph{i.e.,} the distribution and correlations of synaptic weights)
of a naturally occurring network is often strongly influenced by activity
during its formation and maturation. However, studies of network dynamics
typically focus on behavior of the network under given, static network
topology or weight conformation. Advances in the understanding of
the network plasticity are setting the stage for addressing how the
observed conformation of these networks can come about. Our study
of a pure and simple network is just an initial step along this direction,
and it leaves open questions of how variations of biological details
that are fixed in our model and different network topology, geometry,
or sparseness might affect the emergent structures and alter the dynamical
behavior of the network.

In addition, there remain pronounced and puzzling features observed
in the current study that require further elucidation. For example,
while both types of conformations (hubs and paths) appear abruptly
as the plasticity parameter $w^{\star}$ is varied, the average firing
rate increases with a discontinuous jump for the path conformation,
but continuously for the hub. Also, the system does not {}``morph''
from the path conformation to the hub conformation directly but, instead,
returns to a uniform network before developing the hub structure.
This feature might suggest a symmetry or {}``topological'' difference
between the two types of conformations that prevents a direct transition
from one conformation to the other.

Finally, while we have only considered a stationary network driven
by uncorrelated noise, arguably the ultimate {}``goal'' of a plastic
neural network is learning to serve specific or varying functional
requirements. On this regard, it will be of great interest to subject
the system to more meaningful inputs and explore the change in the
stationary structures as well as the transient dynamics of the resultant
network.
\begin{acknowledgments}
This research was supported in part by Computational Resources on
PittGrid (www.pittgrid.pitt.edu).
\end{acknowledgments}
\bibliographystyle{apsrev4-1}

\begin{thebibliography}{10}%
\makeatletter
\providecommand \@ifxundefined [1]{%
 \ifx #1\undefined \expandafter \@firstoftwo
 \else \expandafter \@secondoftwo
\fi
}%
\providecommand \@ifnum [1]{%
 \ifnum #1\expandafter \@firstoftwo
 \else \expandafter \@secondoftwo
\fi
}%
\providecommand \enquote [1]{``#1''}%
\providecommand \bibnamefont  [1]{#1}%
\providecommand \bibfnamefont [1]{#1}%
\providecommand \citenamefont [1]{#1}%
\providecommand\href[0]{\@sanitize\@href}%
\providecommand\@href[1]{\endgroup\@@startlink{#1}\endgroup\@@href}%
\providecommand\@@href[1]{#1\@@endlink}%
\providecommand \@sanitize [0]{\begingroup\catcode`\&12\catcode`\#12\relax}%
\@ifxundefined \pdfoutput {\@firstoftwo}{%
 \@ifnum{\z@=\pdfoutput}{\@firstoftwo}{\@secondoftwo}%
}{%
 \providecommand\@@startlink[1]{\leavevmode}%
 \providecommand\@@endlink[0]{}%
}{%
 \providecommand\@@startlink[1]{%
  \leavevmode
  \pdfstartlink
   attr{/Border[0 0 1 ]/H/I/C[0 1 1]}%
   user{/Subtype/Link/A<</Type/Action/S/URI/URI(#1)>>}%
  \relax
 }%
 \providecommand\@@endlink[0]{\pdfendlink}%
}%
\providecommand \url  [0]{\begingroup\@sanitize \@url }%
\providecommand \@url [1]{\endgroup\@href {#1}{\urlprefix}}%
\providecommand \urlprefix [0]{URL }%
\providecommand \Eprint[0]{\href }%
\@ifxundefined \urlstyle {%
  \providecommand \doi [1]{doi:\discretionary{}{}{}#1}%
}{%
  \providecommand \doi [0]{doi:\discretionary{}{}{}\begingroup
  \urlstyle{rm}\Url }%
}%
\providecommand \doibase [0]{http://dx.doi.org/}%
\providecommand \Doi[1]{\href{\doibase#1}}%
\providecommand \bibAnnote [3]{%
  \BibitemShut{#1}%
  \begin{quotation}\noindent
    \textsc{Key:}\ #2\\\textsc{Annotation:}\ #3%
  \end{quotation}%
}%
\providecommand \bibAnnoteFile [2]{%
  \IfFileExists{#2}{\bibAnnote {#1} {#2} {\input{#2}}}{}%
}%
\providecommand \typeout [0]{\immediate \write \m@ne }%
\providecommand \selectlanguage [0]{\@gobble}%
\providecommand \bibinfo [0]{\@secondoftwo}%
\providecommand \bibfield [0]{\@secondoftwo}%
\providecommand \translation [1]{[#1]}%
\providecommand \BibitemOpen[0]{}%
\providecommand \bibitemStop [0]{}%
\providecommand \bibitemNoStop [0]{.\EOS\space}%
\providecommand \EOS [0]{\spacefactor3000\relax}%
\providecommand \BibitemShut [1]{\csname bibitem#1\endcsname}%
\bibitem{abbott_synaptic_2000}%
  \BibitemOpen
  \bibfield{author}{%
  \bibinfo {author} {\bibfnamefont{L.~F.}\ \bibnamefont{Abbott}}\ and\ \bibinfo
  {author} {\bibfnamefont{S.~B.}\ \bibnamefont{Nelson}},\ }%
  \bibfield{journal}{%
  \Doi{10.1038/81453}{\bibinfo {journal} {Nature Neuroscience}}\ }%
  \textbf{\bibinfo {volume} {3}},\ \bibinfo {pages} {1178} (\bibinfo {year}
  {2000})%
  \bibAnnoteFile{NoStop}{abbott_synaptic_2000}%
\bibitem{van_rossum_stable_2000}%
  \BibitemOpen
  \bibfield{author}{%
  \bibinfo {author} {\bibfnamefont{M.~C.~W.}\ \bibnamefont{van Rossum}},
  \bibinfo {author} {\bibfnamefont{G.~Q.}\ \bibnamefont{Bi}},\ and\ \bibinfo
  {author} {\bibfnamefont{G.~G.}\ \bibnamefont{Turrigiano}},\ }%
  \bibfield{journal}{%
  \bibinfo {journal} {Journal of Neuroscience}\ }%
  \textbf{\bibinfo {volume} {20}},\ \bibinfo {pages} {8812} (\bibinfo {month}
  {Dec.}\ \bibinfo {year} {2000})%
  \bibAnnoteFile{NoStop}{van_rossum_stable_2000}%
\bibitem{roberts_spike_2002}%
  \BibitemOpen
  \bibfield{author}{%
  \bibinfo {author} {\bibfnamefont{P.~D.}\ \bibnamefont{Roberts}}\ and\
  \bibinfo {author} {\bibfnamefont{C.~C.}\ \bibnamefont{Bell}},\ }%
  \bibfield{journal}{%
  \Doi{10.1007/s00422-002-0361-y}{\bibinfo {journal} {Biological Cybernetics}}\
  }%
  \textbf{\bibinfo {volume} {87}},\ \bibinfo {pages} {392} (\bibinfo {month}
  {Dec.}\ \bibinfo {year} {2002})%
  \bibAnnoteFile{NoStop}{roberts_spike_2002}%
\bibitem{dan_spike_2004}%
  \BibitemOpen
  \bibfield{author}{%
  \bibinfo {author} {\bibfnamefont{Y.}~\bibnamefont{Dan}}\ and\ \bibinfo
  {author} {\bibfnamefont{M.-M.}\ \bibnamefont{Poo}},\ }%
  \bibfield{journal}{%
  \Doi{10.1016/j.neuron.2004.09.007}{\bibinfo {journal} {Neuron}}\ }%
  \textbf{\bibinfo {volume} {44}},\ \bibinfo {pages} {23} (\bibinfo {month}
  {Sep.}\ \bibinfo {year} {2004}),\ ISSN \bibinfo {issn} {0896-6273}%
  \bibAnnoteFile{NoStop}{dan_spike_2004}%
\bibitem{morrison_spike-timing-dependent_2007}%
  \BibitemOpen
  \bibfield{author}{%
  \bibinfo {author} {\bibfnamefont{A.}~\bibnamefont{Morrison}}, \bibinfo
  {author} {\bibfnamefont{A.}~\bibnamefont{Aertsen}},\ and\ \bibinfo {author}
  {\bibfnamefont{M.}~\bibnamefont{Diesmann}},\ }%
  \bibfield{journal}{%
  \Doi{10.1162/neco.2007.19.6.1437}{\bibinfo {journal} {Neural Computation}}\
  }%
  \textbf{\bibinfo {volume} {19}},\ \bibinfo {pages} {1437} (\bibinfo {month}
  {Jun.}\ \bibinfo {year} {2007})%
  \bibAnnoteFile{NoStop}{morrison_spike-timing-dependent_2007}%
\bibitem{chen_mean-field_2010}%
  \BibitemOpen
  \bibfield{author}{%
  \bibinfo {author} {\bibfnamefont{C.-C.}\ \bibnamefont{Chen}}\ and\ \bibinfo
  {author} {\bibfnamefont{D.}~\bibnamefont{Jasnow}},\ }%
  \bibfield{journal}{%
  \Doi{10.1103/PhysRevE.81.011907}{\bibinfo {journal} {Physical Review E}}\ }%
  \textbf{\bibinfo {volume} {81}},\ \bibinfo {pages} {011907} (\bibinfo {year}
  {2010})%
  \bibAnnoteFile{NoStop}{chen_mean-field_2010}%
\bibitem{gerstner_spiking_2002}%
  \BibitemOpen
  \bibfield{author}{%
  \bibinfo {author} {\bibfnamefont{W.}~\bibnamefont{Gerstner}}\ and\ \bibinfo
  {author} {\bibfnamefont{W.~M.}\ \bibnamefont{Kistler}},\ }%
  \emph{\bibinfo {title} {Spiking Neuron Models: Single Neurons, Populations,
  Plasticity}}\ (\bibinfo {publisher} {Cambridge University Press},\ \bibinfo
  {year} {2002})\ ISBN \bibinfo {isbn} {0521890799}%
  \bibAnnoteFile{NoStop}{gerstner_spiking_2002}%
\bibitem{burkitt_review_2006}%
  \BibitemOpen
  \bibfield{author}{%
  \bibinfo {author} {\bibfnamefont{A.}~\bibnamefont{Burkitt}},\ }%
  \bibfield{journal}{%
  \Doi{10.1007/s00422-006-0082-8}{\bibinfo {journal} {Biological Cybernetics}}\
  }%
  \textbf{\bibinfo {volume} {95}},\ \bibinfo {pages} {97} (\bibinfo {year}
  {2006})%
  \bibAnnoteFile{NoStop}{burkitt_review_2006}%
\bibitem{brette_exact_2006}%
  \BibitemOpen
  \bibfield{author}{%
  \bibinfo {author} {\bibfnamefont{R.}~\bibnamefont{Brette}},\ }%
  \bibfield{journal}{%
  \Doi{10.1162/neco.2006.18.8.2004}{\bibinfo {journal} {Neural Computation}}\
  }%
  \textbf{\bibinfo {volume} {18}},\ \bibinfo {pages} {2004} (\bibinfo {year}
  {2006})%
  \bibAnnoteFile{NoStop}{brette_exact_2006}%
\bibitem{tsodyks_synchrony_2000}%
  \BibitemOpen
  \bibfield{author}{%
  \bibinfo {author} {\bibfnamefont{M.}~\bibnamefont{Tsodyks}}, \bibinfo
  {author} {\bibfnamefont{A.}~\bibnamefont{Uziel}},\ and\ \bibinfo {author}
  {\bibfnamefont{H.}~\bibnamefont{Markram}},\ }%
  \bibfield{journal}{%
  \bibinfo {journal} {Journal of Neuroscience}\ }%
  \textbf{\bibinfo {volume} {20}},\ \bibinfo {pages} {RC50} (\bibinfo {year}
  {2000})%
  \bibAnnoteFile{NoStop}{tsodyks_synchrony_2000}%
\bibitem{brette_simulation_2007}%
  \BibitemOpen
  \bibfield{author}{%
  \bibinfo {author} {\bibfnamefont{R.}~\bibnamefont{Brette}}, \bibinfo {author}
  {\bibfnamefont{M.}~\bibnamefont{Rudolph}}, \bibinfo {author}
  {\bibfnamefont{T.}~\bibnamefont{Carnevale}}, \bibinfo {author}
  {\bibfnamefont{M.}~\bibnamefont{Hines}}, \bibinfo {author}
  {\bibfnamefont{D.}~\bibnamefont{Beeman}}, \bibinfo {author}
  {\bibfnamefont{J.}~\bibnamefont{Bower}}, \bibinfo {author}
  {\bibfnamefont{M.}~\bibnamefont{Diesmann}}, \bibinfo {author}
  {\bibfnamefont{A.}~\bibnamefont{Morrison}}, \bibinfo {author}
  {\bibfnamefont{P.}~\bibnamefont{Goodman}}, \bibinfo {author}
  {\bibfnamefont{F.}~\bibnamefont{Harris}}, \bibinfo {author}
  {\bibfnamefont{M.}~\bibnamefont{Zirpe}}, \bibinfo {author}
  {\bibfnamefont{T.}~\bibnamefont{Natschl\"ager}}, \bibinfo {author}
  {\bibfnamefont{D.}~\bibnamefont{Pecevski}}, \bibinfo {author}
  {\bibfnamefont{B.}~\bibnamefont{Ermentrout}}, \bibinfo {author}
  {\bibfnamefont{M.}~\bibnamefont{Djurfeldt}}, \bibinfo {author}
  {\bibfnamefont{A.}~\bibnamefont{Lansner}}, \bibinfo {author}
  {\bibfnamefont{O.}~\bibnamefont{Rochel}}, \bibinfo {author}
  {\bibfnamefont{T.}~\bibnamefont{Vieville}}, \bibinfo {author}
  {\bibfnamefont{E.}~\bibnamefont{Muller}}, \bibinfo {author}
  {\bibfnamefont{A.}~\bibnamefont{Davison}}, \bibinfo {author}
  {\bibfnamefont{S.~E.}\ \bibnamefont{Boustani}},\ and\ \bibinfo {author}
  {\bibfnamefont{A.}~\bibnamefont{Destexhe}},\ }%
  \bibfield{journal}{%
  \Doi{10.1007/s10827-007-0038-6}{\bibinfo {journal} {Journal of Computational
  Neuroscience}}\ }%
  \textbf{\bibinfo {volume} {23}},\ \bibinfo {pages} {349} (\bibinfo {month}
  {Dec.}\ \bibinfo {year} {2007})%
  \bibAnnoteFile{NoStop}{brette_simulation_2007}%
\bibitem{dayan_theoretical_2001}%
  \BibitemOpen
  \bibfield{author}{%
  \bibinfo {author} {\bibfnamefont{P.}~\bibnamefont{Dayan}}\ and\ \bibinfo
  {author} {\bibfnamefont{L.~F.}\ \bibnamefont{Abbott}},\ }%
  \emph{\bibinfo {title} {Theoretical Neuroscience: Computational and
  Mathematical Modeling of Neural Systems}}\ (\bibinfo {publisher}
  {Massachusetts Institute of Technology Press},\ \bibinfo {address}
  {Cambridge, Mass},\ \bibinfo {year} {2001})\ ISBN \bibinfo {isbn}
  {0262041995}%
  \bibAnnoteFile{NoStop}{dayan_theoretical_2001}%
\bibitem{amarasingham_predictingdistribution_1998}%
  \BibitemOpen
  \bibfield{author}{%
  \bibinfo {author} {\bibfnamefont{A.}~\bibnamefont{Amarasingham}}\ and\
  \bibinfo {author} {\bibfnamefont{W.~B.}\ \bibnamefont{Levy}},\ }%
  \bibfield{journal}{%
  \Doi{10.1162/089976698300017881}{\bibinfo {journal} {Neural Computation}}\ }%
  \textbf{\bibinfo {volume} {10}},\ \bibinfo {pages} {25} (\bibinfo {year}
  {1998})%
  \bibAnnoteFile{NoStop}{amarasingham_predictingdistribution_1998}%
\bibitem{song_competitive_2000}%
  \BibitemOpen
  \bibfield{author}{%
  \bibinfo {author} {\bibfnamefont{S.}~\bibnamefont{Song}}, \bibinfo {author}
  {\bibfnamefont{K.~D.}\ \bibnamefont{Miller}},\ and\ \bibinfo {author}
  {\bibfnamefont{L.~F.}\ \bibnamefont{Abbott}},\ }%
  \bibfield{journal}{%
  \Doi{10.1038/78829}{\bibinfo {journal} {Nature Neuroscience}}\ }%
  \textbf{\bibinfo {volume} {3}},\ \bibinfo {pages} {919} (\bibinfo {year}
  {2000}),\ ISSN \bibinfo {issn} {1097-6256}%
  \bibAnnoteFile{NoStop}{song_competitive_2000}%
\bibitem{rubin_equilibrium_2001}%
  \BibitemOpen
  \bibfield{author}{%
  \bibinfo {author} {\bibfnamefont{J.}~\bibnamefont{Rubin}}, \bibinfo {author}
  {\bibfnamefont{D.~D.}\ \bibnamefont{Lee}},\ and\ \bibinfo {author}
  {\bibfnamefont{H.}~\bibnamefont{Sompolinsky}},\ }%
  \bibfield{journal}{%
  \Doi{10.1103/PhysRevLett.86.364}{\bibinfo {journal} {Physical Review
  Letters}}\ }%
  \textbf{\bibinfo {volume} {86}},\ \bibinfo {pages} {364} (\bibinfo {year}
  {2001})%
  \bibAnnoteFile{NoStop}{rubin_equilibrium_2001}%
\bibitem{toyoizumi_optimality_2007}%
  \BibitemOpen
  \bibfield{author}{%
  \bibinfo {author} {\bibfnamefont{T.}~\bibnamefont{Toyoizumi}}, \bibinfo
  {author} {\bibfnamefont{J.-P.}\ \bibnamefont{Pfister}}, \bibinfo {author}
  {\bibfnamefont{K.}~\bibnamefont{Aihara}},\ and\ \bibinfo {author}
  {\bibfnamefont{W.}~\bibnamefont{Gerstner}},\ }%
  \bibfield{journal}{%
  \Doi{10.1162/neco.2007.19.3.639}{\bibinfo {journal} {Neural Computation}}\ }%
  \textbf{\bibinfo {volume} {19}},\ \bibinfo {pages} {639} (\bibinfo {month}
  {Mar.}\ \bibinfo {year} {2007})%
  \bibAnnoteFile{NoStop}{toyoizumi_optimality_2007}%
\bibitem{hasselmo_runaway_1994}%
  \BibitemOpen
  \bibfield{author}{%
  \bibinfo {author} {\bibfnamefont{M.~E.}\ \bibnamefont{Hasselmo}},\ }%
  \bibfield{journal}{%
  \Doi{10.1016/0893-6080(94)90053-1}{\bibinfo {journal} {Neural Networks}}\ }%
  \textbf{\bibinfo {volume} {7}},\ \bibinfo {pages} {13} (\bibinfo {year}
  {1994}),\ ISSN \bibinfo {issn} {0893-6080}%
  \bibAnnoteFile{NoStop}{hasselmo_runaway_1994}%
\bibitem{greenstein-messica_synaptic_2011}%
  \BibitemOpen
  \bibfield{author}{%
  \bibinfo {author} {\bibfnamefont{A.}~\bibnamefont{Greenstein-Messica}}\ and\
  \bibinfo {author} {\bibfnamefont{E.}~\bibnamefont{Ruppin}},\ }%
  \bibfield{journal}{%
  \Doi{10.1162/089976698300017836}{\bibinfo {journal} {Neural Computation}}\ }%
  \textbf{\bibinfo {volume} {10}},\ \bibinfo {pages} {451} (\bibinfo {year}
  {2011}),\ ISSN \bibinfo {issn} {0899-7667}%
  \bibAnnoteFile{NoStop}{greenstein-messica_synaptic_2011}%
\bibitem{bullmore_complex_2009}%
  \BibitemOpen
  \bibfield{author}{%
  \bibinfo {author} {\bibfnamefont{E.}~\bibnamefont{Bullmore}}\ and\ \bibinfo
  {author} {\bibfnamefont{O.}~\bibnamefont{Sporns}},\ }%
  \bibfield{journal}{%
  \Doi{10.1038/nrn2575}{\bibinfo {journal} {Nature Reviews Neuroscience}}\ }%
  \textbf{\bibinfo {volume} {10}},\ \bibinfo {pages} {186} (\bibinfo {month}
  {Mar.}\ \bibinfo {year} {2009}),\ ISSN \bibinfo {issn} {1471-003X}%
  \bibAnnoteFile{NoStop}{bullmore_complex_2009}%
\bibitem{boccaletti_complex_2006}%
  \BibitemOpen
  \bibfield{author}{%
  \bibinfo {author} {\bibfnamefont{S.}~\bibnamefont{Boccaletti}}, \bibinfo
  {author} {\bibfnamefont{V.}~\bibnamefont{Latora}}, \bibinfo {author}
  {\bibfnamefont{Y.}~\bibnamefont{Moreno}}, \bibinfo {author}
  {\bibfnamefont{M.}~\bibnamefont{Chavez}},\ and\ \bibinfo {author}
  {\bibfnamefont{D.-U.}\ \bibnamefont{Hwang}},\ }%
  \bibfield{journal}{%
  \Doi{10.1016/j.physrep.2005.10.009}{\bibinfo {journal} {Physics Reports}}\ }%
  \textbf{\bibinfo {volume} {424}},\ \bibinfo {pages} {175} (\bibinfo {month}
  {Feb.}\ \bibinfo {year} {2006})%
  \bibAnnoteFile{NoStop}{boccaletti_complex_2006}%
\bibitem{newman_modularity_2006}%
  \BibitemOpen
  \bibfield{author}{%
  \bibinfo {author} {\bibfnamefont{M.~E.~J.}\ \bibnamefont{Newman}},\ }%
  \bibfield{journal}{%
  \Doi{10.1073/pnas.0601602103}{\bibinfo {journal} {Proceedings of the National
  Academy of Sciences}}\ }%
  \textbf{\bibinfo {volume} {103}},\ \bibinfo {pages} {8577 } (\bibinfo {month}
  {Jun.}\ \bibinfo {year} {2006})%
  \bibAnnoteFile{NoStop}{newman_modularity_2006}%
\bibitem{Note1}%
  \BibitemOpen
  \bibinfo {note} {The loop product $w_{i,j}w_{j,k}w_{k,i}$ is an attempt to
  generalize the curl operator $\nabla \times $ in metric spaces to a network.
  The condition \protect \textup {\hbox {\mathsurround \z@ \protect \normalfont
  (\ignorespaces \ref {eq:triplet-weight}\unskip \@@italiccorr )}} is
  equivalent to the curl-free condition of the {}``vector field''
  $v_{i,j}\equiv \protect \qopname \relax o{ln}\left (w_{i,j}/w^{\star }\right
  )$, which seems to hold in the hub conformations but not the loop or path
  conformations for the stationary plastic networks.}%
  \bibAnnoteFile{Stop}{Note1}%
\bibitem{bi_synaptic_1998}%
  \BibitemOpen
  \bibfield{author}{%
  \bibinfo {author} {\bibfnamefont{G.-Q.}\ \bibnamefont{Bi}}\ and\ \bibinfo
  {author} {\bibfnamefont{M.-M.}\ \bibnamefont{Poo}},\ }%
  \bibfield{journal}{%
  \bibinfo {journal} {Journal of Neuroscience}\ }%
  \textbf{\bibinfo {volume} {18}},\ \bibinfo {pages} {10464} (\bibinfo {month}
  {Dec.}\ \bibinfo {year} {1998})%
  \bibAnnoteFile{NoStop}{bi_synaptic_1998}%
\bibitem{segev_observations_2001}%
  \BibitemOpen
  \bibfield{author}{%
  \bibinfo {author} {\bibfnamefont{R.}~\bibnamefont{Segev}}, \bibinfo {author}
  {\bibfnamefont{Y.}~\bibnamefont{Shapira}}, \bibinfo {author}
  {\bibfnamefont{M.}~\bibnamefont{Benveniste}},\ and\ \bibinfo {author}
  {\bibfnamefont{E.}~\bibnamefont{Ben-Jacob}},\ }%
  \bibfield{journal}{%
  \bibinfo {journal} {Physical Review E}\ }%
  \textbf{\bibinfo {volume} {64}},\ \bibinfo {pages} {011920} (\bibinfo {month}
  {Jun.}\ \bibinfo {year} {2001})%
  \bibAnnoteFile{NoStop}{segev_observations_2001}%
\bibitem{shefi_morphological_2002}%
  \BibitemOpen
  \bibfield{author}{%
  \bibinfo {author} {\bibfnamefont{O.}~\bibnamefont{Shefi}}, \bibinfo {author}
  {\bibfnamefont{I.}~\bibnamefont{Golding}}, \bibinfo {author}
  {\bibfnamefont{R.}~\bibnamefont{Segev}}, \bibinfo {author}
  {\bibfnamefont{E.}~\bibnamefont{Ben-Jacob}},\ and\ \bibinfo {author}
  {\bibfnamefont{A.}~\bibnamefont{Ayali}},\ }%
  \bibfield{journal}{%
  \bibinfo {journal} {Physical Review E}\ }%
  \textbf{\bibinfo {volume} {66}},\ \bibinfo {pages} {021905} (\bibinfo {year}
  {2002})%
  \bibAnnoteFile{NoStop}{shefi_morphological_2002}%
\bibitem{beggs_neuronal_2003}%
  \BibitemOpen
  \bibfield{author}{%
  \bibinfo {author} {\bibfnamefont{J.~M.}\ \bibnamefont{Beggs}}\ and\ \bibinfo
  {author} {\bibfnamefont{D.}~\bibnamefont{Plenz}},\ }%
  \bibfield{journal}{%
  \bibinfo {journal} {J. Neurosci.}\ }%
  \textbf{\bibinfo {volume} {23}},\ \bibinfo {pages} {11167} (\bibinfo {month}
  {Dec.}\ \bibinfo {year} {2003})%
  \bibAnnoteFile{NoStop}{beggs_neuronal_2003}%
\bibitem{lai_growth_2006}%
  \BibitemOpen
  \bibfield{author}{%
  \bibinfo {author} {\bibfnamefont{P.-Y.}\ \bibnamefont{Lai}}, \bibinfo
  {author} {\bibfnamefont{L.~C.}\ \bibnamefont{Jia}},\ and\ \bibinfo {author}
  {\bibfnamefont{C.~K.}\ \bibnamefont{Chan}},\ }%
  \bibfield{journal}{%
  \bibinfo {journal} {Physical Review E}\ }%
  \textbf{\bibinfo {volume} {73}},\ \bibinfo {pages} {051906} (\bibinfo {month}
  {May}\ \bibinfo {year} {2006})%
  \bibAnnoteFile{NoStop}{lai_growth_2006}%
\bibitem{barrat_architecture_2004}%
  \BibitemOpen
  \bibfield{author}{%
  \bibinfo {author} {\bibfnamefont{A.}~\bibnamefont{Barrat}}, \bibinfo {author}
  {\bibfnamefont{M.}~\bibnamefont{Barthelemy}}, \bibinfo {author}
  {\bibfnamefont{R.}~\bibnamefont{Pastor-Satorras}},\ and\ \bibinfo {author}
  {\bibfnamefont{A.}~\bibnamefont{Vespignani}},\ }%
  \bibfield{journal}{%
  \Doi{10.1073/pnas.0400087101}{\bibinfo {journal} {Proceedings of the National
  Academy of Sciences of the United States of America}}\ }%
  \textbf{\bibinfo {volume} {101}},\ \bibinfo {pages} {3747} (\bibinfo {year}
  {2004})%
  \bibAnnoteFile{NoStop}{barrat_architecture_2004}%
\end{thebibliography}
\end{document}